\documentclass{PoS}

\usepackage{array}
 \usepackage{multicol}

\title{Production of W+jets in Relativistic heavy-ion collisions}

\ShortTitle{W+jet correlations in heavy-ion collisions}

\author{\speaker{Shan-Liang Zhang} $^a$, Xin-Nian Wang$^{ab}$, Ben-Wei Zhang$^a$  \\
        \llap{$^a$} Institute of Particle Physics and Key Laboratory of Quarks and Lepton Physics (MOE), Central China Normal University, Wuhan 430079, China\\
        \llap{$^b$} Nuclear Science Division Mailstop 70R0319, Lawrence Berkeley National Laboratory, Berkeley, CA 94740\\
        E-mail: \email{zhangshanl@mails.ccnu.edu.cn},  \email{xnwang@lbl.gov}, \email{bwzhang@mail.ccnu.edu.cn}}

\abstract{We carry out a  detailed calculations of W+jets production in Pb+Pb collisions at the Large Hadron Collider (LHC). In our calculations, the production of W+jet in p+p reference is obtained from Sherpa, which matches next-to-leading-order matrix elements to the resummation of parton shower calculations. Jet propagation and medium response in the quark-gluon plasma is simulated with the Linear Boltzmann Transport (LBT) model. We calculate five observables
of W+jets productions with jet quenching effect in Pb+Pb collisions: event distribution as a function of the vector sum of the lepton and jets  $|\vec{p}_T^{Miss}|$,  nuclear effects for tagged jet cross sections $I_{AA}$,  azimuthal angle correlations $\Delta \phi_{jW}$, mean value of momentum imbalance $\langle x_{jW}\rangle$, average number of jets per W boson $R_{jW}$. The  nuclear modifications of these 5 observables due to jet quenching in Pb+Pb relative to that in p+p collisions are discussed.}

\FullConference{%
  HardProbes2020\\
  1-6 June 2020\\
  Austin, Texas}

\begin{document}

\section{Introduction}
 Vector gauge boson associated with jet production are golden channel to probe the properties of the quark-gluon plasma (QGP)~\cite{Neufeld:2010fj}.
Jets energies are reduced due to elastic and inelastic scattering with the hot and dense medium when they propagate through the QGP~\cite{Gyulassy:2003mc}, while vector gauge bosons will not participate in the strong-interactions directly, escaping the QGP unmodified. Therefore, the vector gauge boson transverse momentum closely reflects the initial energy of the associated jet before interacting with the hot-dense medium.

Recently,$\gamma+$jet correlations~\cite{Dai:2012am}, Z+jet production~\cite{Zhang:2018urd}, and H+jet processes~\cite{Chen:2020kex} have already been investigated by several theory models and  experimental groups in Pb-Pb collisions at $\sqrt s=5.02$ TeV. However, so far the production of W+jets in heavy-ion collisions has not yet been quantitatively studied. For completeness, we will carry out a detailed calculations of W+jet production in Pb+Pb collisions at the LHC~\cite{Zhang:2020}. As in Z+jets \cite{Zhang:2018urd}, next-to-leading-order (NLO) calculations do not take the resummation of soft/collinear radiation into account and has only limited number of finial particles. Besides, leading-order matrix elements (ME) merged on parton showers(PS) already contain some high-order corrections from both real and virtual contributions. It is short of additional hard or wide-angle radiation from high-order matrix element calculations. Therefore, one needs  improved reference of gauge-boson associated with jets production in p+p collisions to study W+jets correlations in heavy-ion collisions.

\section{Model setup for W+jet in heavy-ion collisions}

\begin{figure}
  \centering
   \includegraphics[scale=0.25]{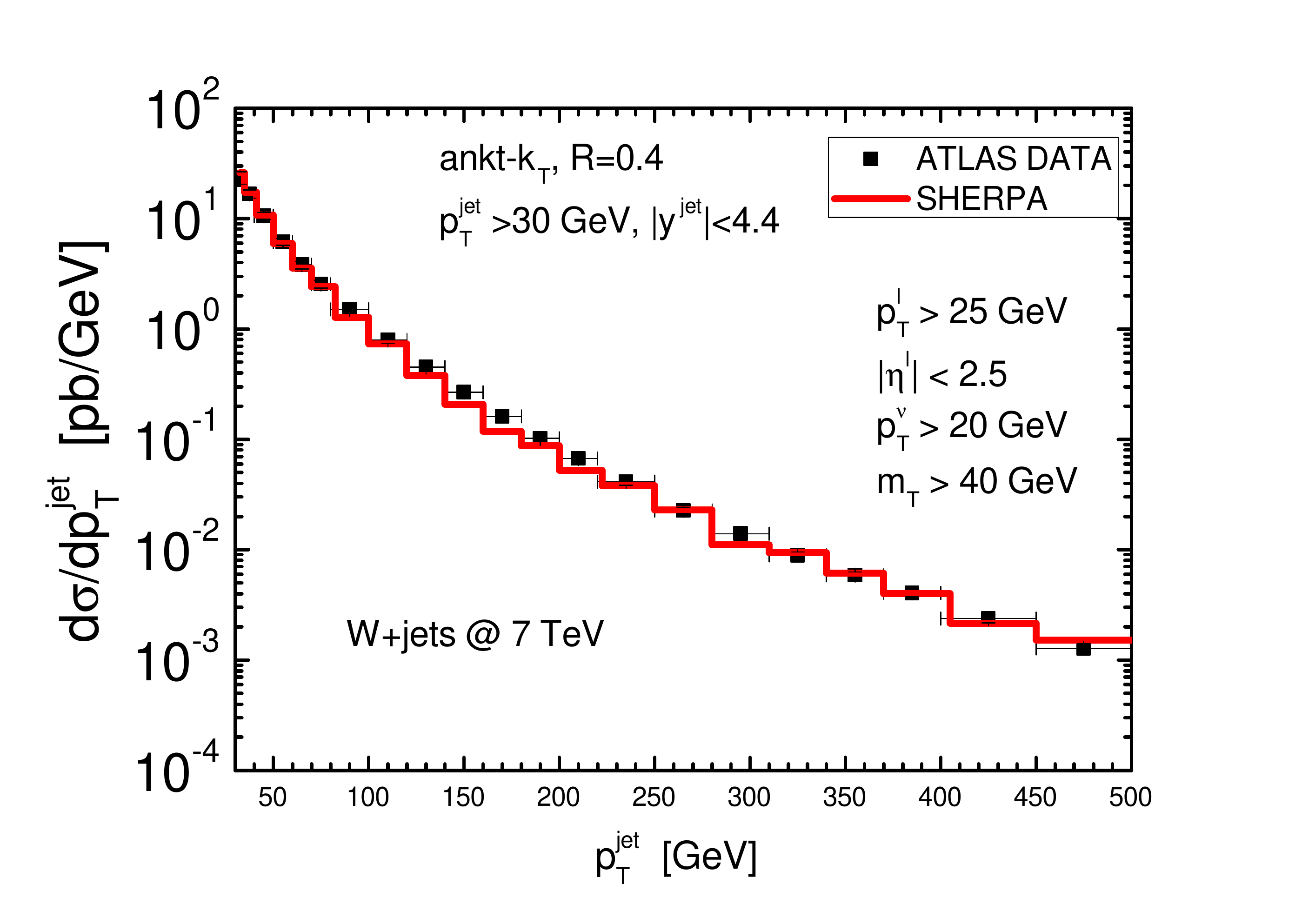}
   \includegraphics[scale=0.25]{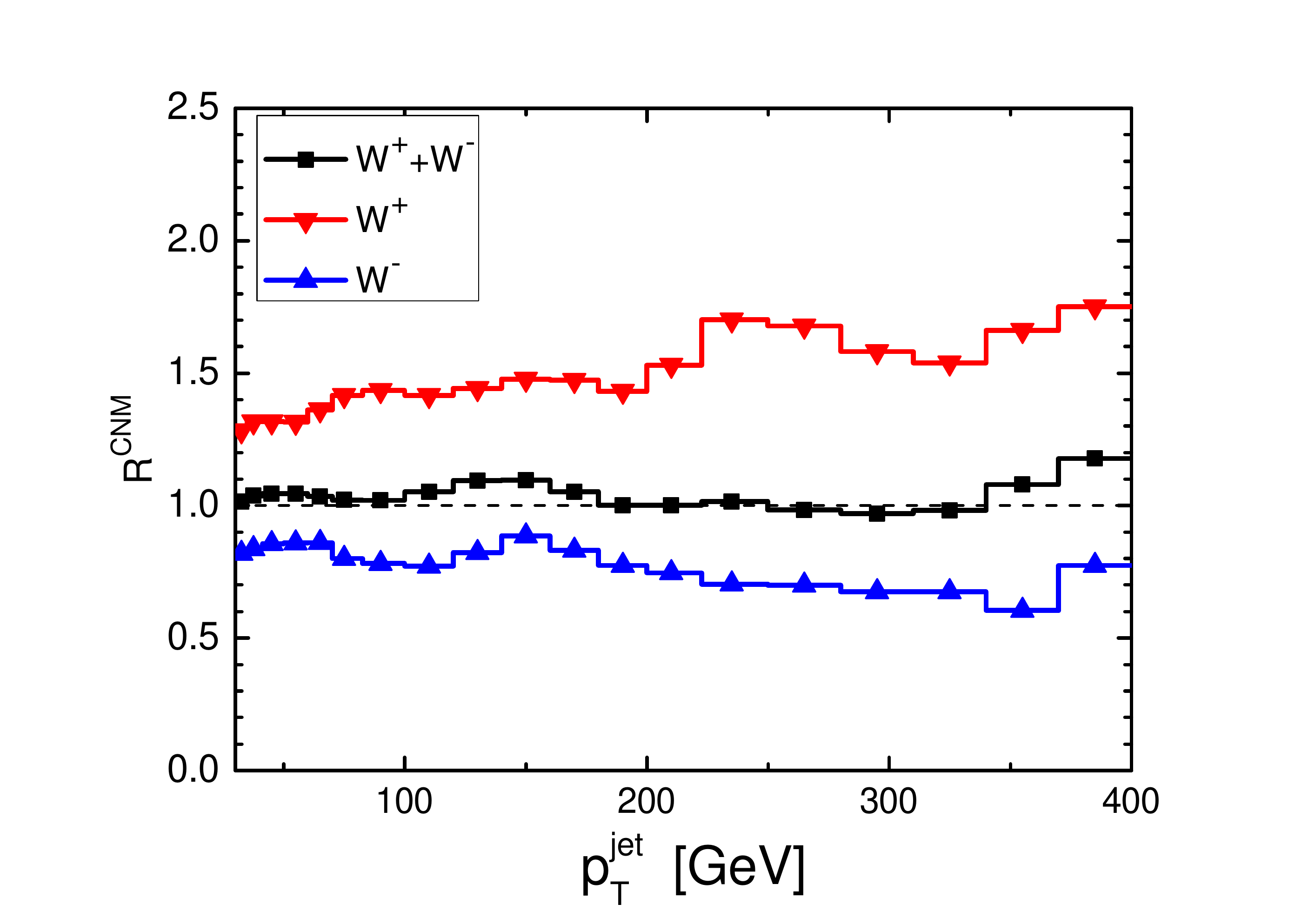}
         \vspace{-10pt}
  \caption{(Color online) Left:Distributions of jet transverse momentum of W+jets in pp collisions at 7 TeV. Right:CNM effects on jet transverse momentum spectra at $\sqrt {s_{NN}}=5.02 $ TeV.
  }\label{baseline}
\end{figure}

 Reference W+jets production in p+p collisions is simulated using NLO matrix elements perturbative calculations matched to the resummation of parton showers~\cite{Hoche:2010kg, Hoeche:2012yf} within the Monte Corlo event generator Sherpa~\cite{Gleisberg:2008ta} at $\sqrt {s_{NN}}=5.02 $ TeV. The differential cross section of W boson associated with jets production as a function of the jet transverse momentum shows well agreement with experimental data~\cite{Aad:2014qxa} as shown in Fig~.\ref{baseline}(left). And then, EPPS16 modified npdfs are used to study cold nuclear matter effects(CNM), as shown in Fig~.\ref{baseline}(right).  The $p_T^{jet}$ spectrum of jets tagged by $W^-$  is significantly suppressed, while the $p_T^{jet}$ spectrum of jets tagged by $W^+$  is significantly enhanced due to CNM effects.  However, the effect of CNM on the total W($W^++W^-$)+jets is consistent with unity and negligible. Our results is in accordance with the calculation in \cite{Ru:2016wfx}. All the physical explanation of the difference of jet spectra for W+jets  between p+p and Pb+Pb collisions should be the result of jet-medium interactions.

The Linear Boltzmann Transport (LBT) model is then used to simulate the propagation, energy attenuation of, and medium
response induced by jet partons in the QGP~\cite{He:2015pra}. LBT is based on a Boltzmann equation ~\cite{He:2015pra}:
\begin{equation}
p_a\cdot\partial f_a(p_a)=-\frac{1}{2}\int\sum _{i=b,c,d}\frac{d^3p_i}{(2\pi)^32E_i}\times[f_af_b-f_cf_d]|M_{ab\rightarrow cd}|^2\times S_2(s,t,u)(2\pi)^4\delta^4(p_a+p_b-p_c-p_d)
\end{equation}

where $f_i$ are phase-space distributions of partons, $S_2(s,t,u)$ is a Lorentz-invariant regulation condition. Elastic scattering is introduced by the complete  set of $2 \rightarrow 2$ matrix elements $|M_{ab\rightarrow cd}|$, and the inelastic scattering is described by the higher-twist formalism for induced gluon radiation ~\cite{Guo:2000nz, Zhang:2003wk, Schafer:2007xh}.

 \section{Numerical results}

 To be consistent with experimental data, W and the associated jets are selected according to the kinematic cut adopted by ATLAS experiment~\cite{Aad:2014qxa}. The information of the evolving bulk matter is provided by (3+1)D hydrodynamics~\cite{Pang:2012he}.

\begin{figure}[htb!]
\centering

  \begin{tabular}{p{6.5cm}<{\centering} p{8cm}<{\centering} }
\includegraphics[width=0.65\textwidth]{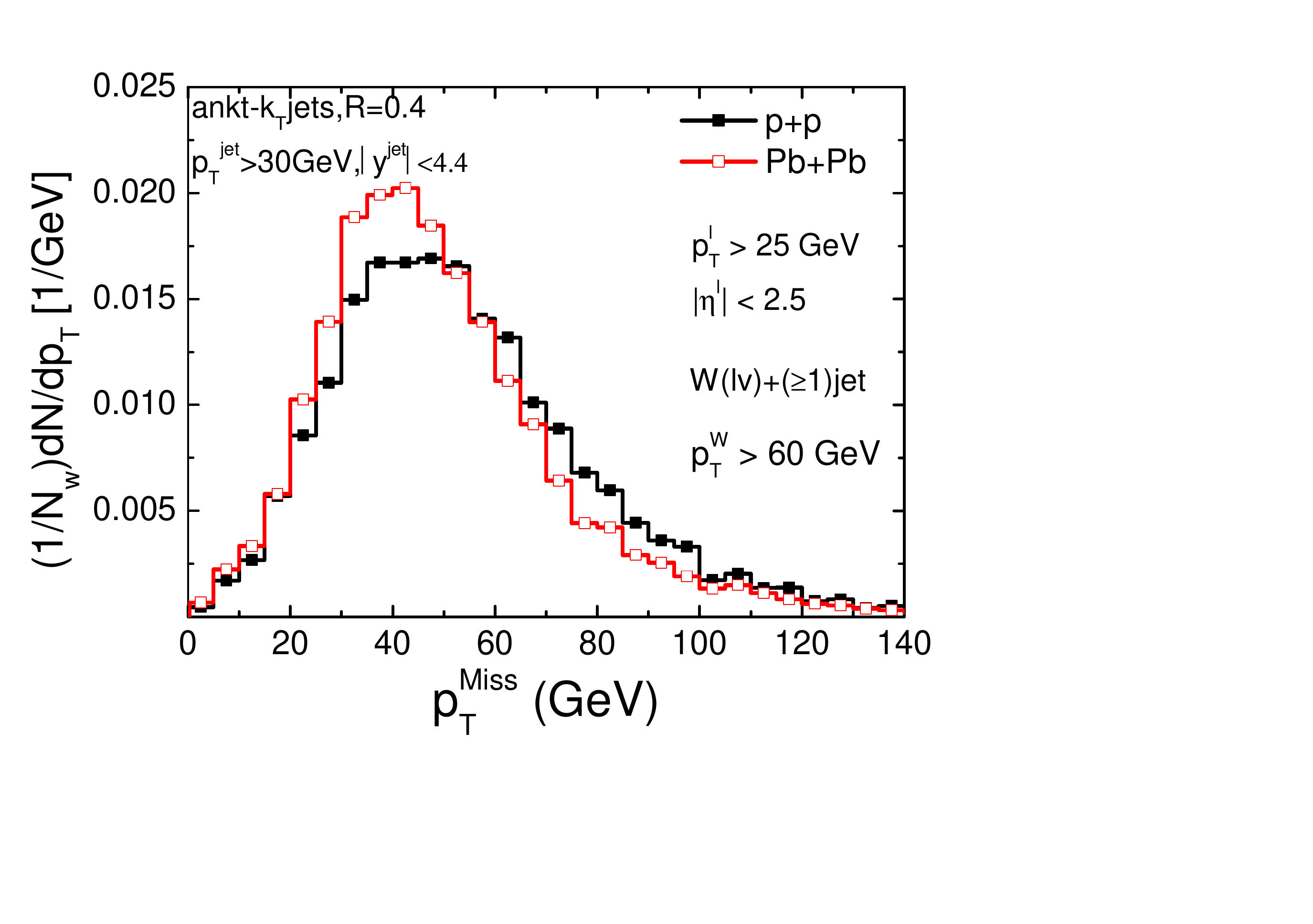} &
\includegraphics[width=0.65\textwidth]{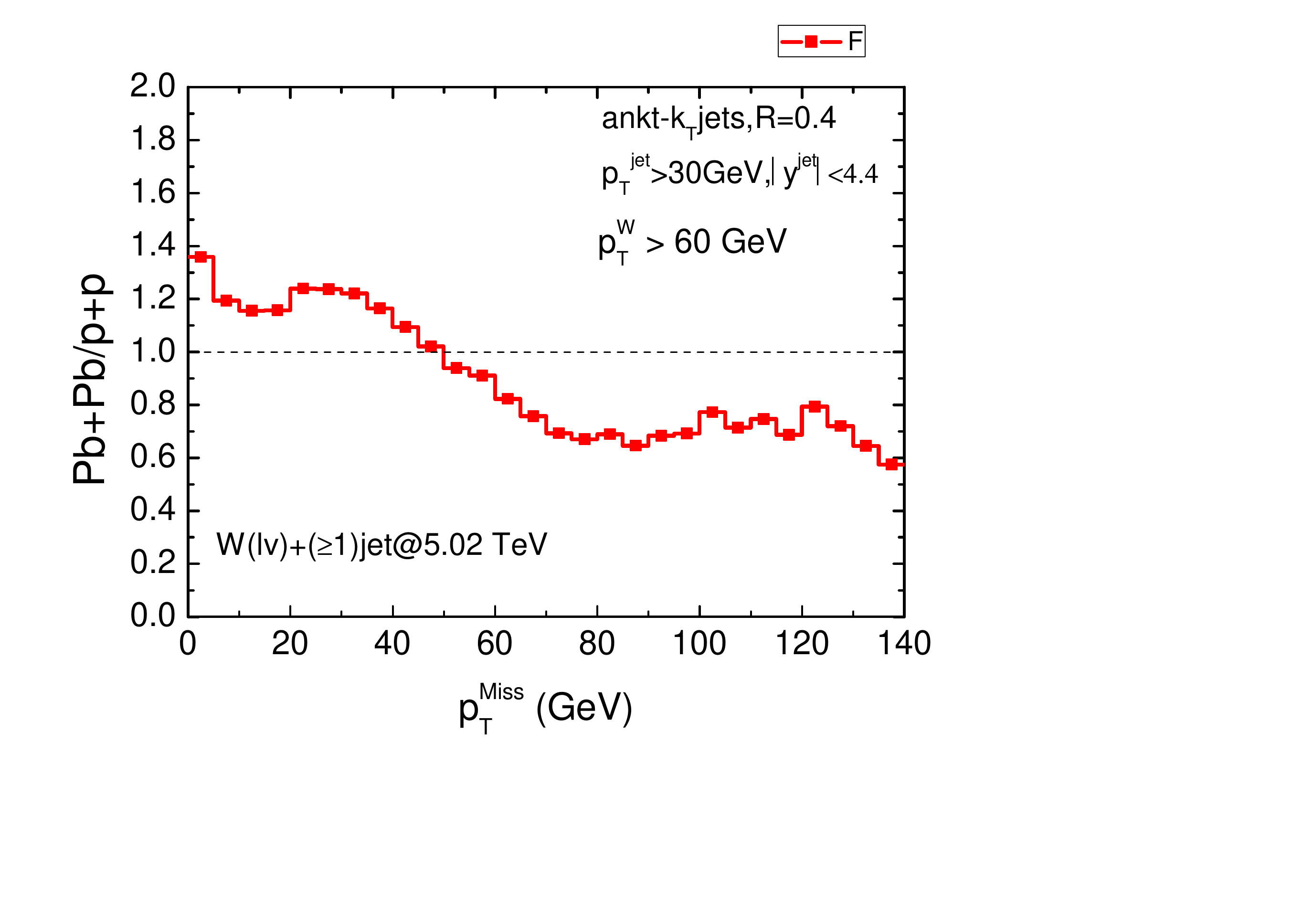}  \\
\end{tabular}
\vspace{-35pt}
\caption{(Color online) Left: Normalized distributions of events passing the W + jets selection cut as a function of the $\vec{p}_T^{Miss}$ . Right:   the ratio of distributions in Pb+Pb to that in p+p collisions at $\sqrt s=5.02$ TeV. }\label{ptmissing}
\end{figure}

Firstly, we present an observable defined  by  the lepton and jets $p_T$ as  $\vec{p}_T^{Miss}=-(\vec{p}_T^{\ l}+ \sum \vec{p}_T^{jets})$.
Since  W boson eventually decays into an electron and a neutrino which is rather difficult to be measured. So the transverse momentum of neutrino is calculated as the negative vector sum of the $p_T$ of electrons, jets, and other soft clusters because of energy-momentum conservation.
So, in p+p collisions, $p_T^{Miss}$ is just similar equal to the transverse momentum of the neutrino. In Pb+Pb collisions, it is the vector sum of
  the transverse momentum that jets radiate out of jet cone and the neutrino transverse momentum. The distributions of $p_T^{Miss}$ from  pp and PbPb collisions are shown in Fig~.\ref{ptmissing}. The distribution of $\vec{p}_T^{Miss}$  is shifted to smaller value in Pb+Pb collisions relative to p+p collisions. This displacement is a result of that some amount of jets transverse energy is radiated out of jet cone due to elastic and inelastic scattering with the medium and the direction of the momentum carried by partons which are radiated by jet out of jet cone is in the opposite direction of the neutrino or W boson. It would simplify the experimental measurements because we just need the jets information which can be easily measured and calculated in W+jets events.

\begin{figure}[tpb]
 \begin{center} \centering

  \includegraphics[scale=0.25]{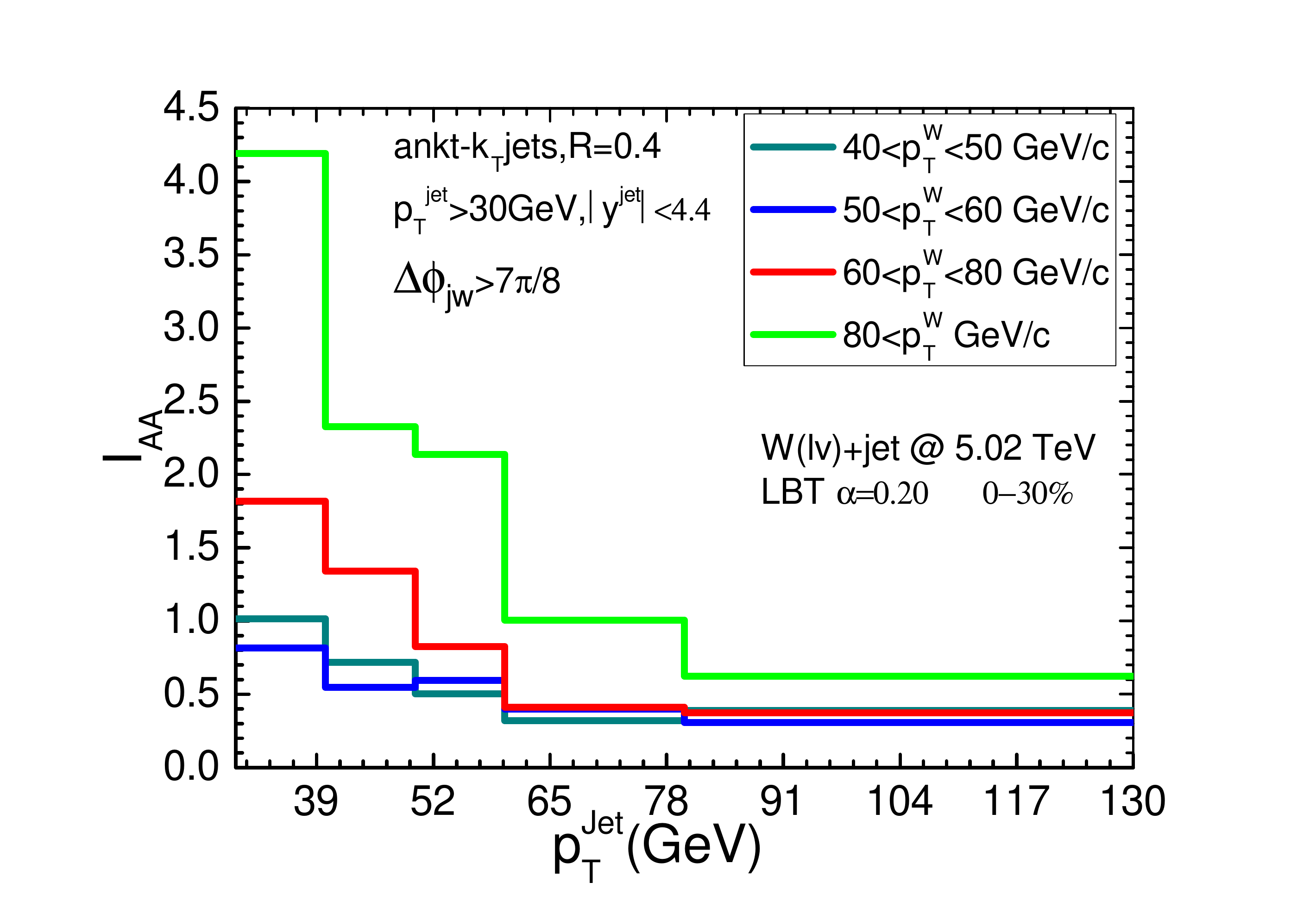}
   \includegraphics[scale=0.25]{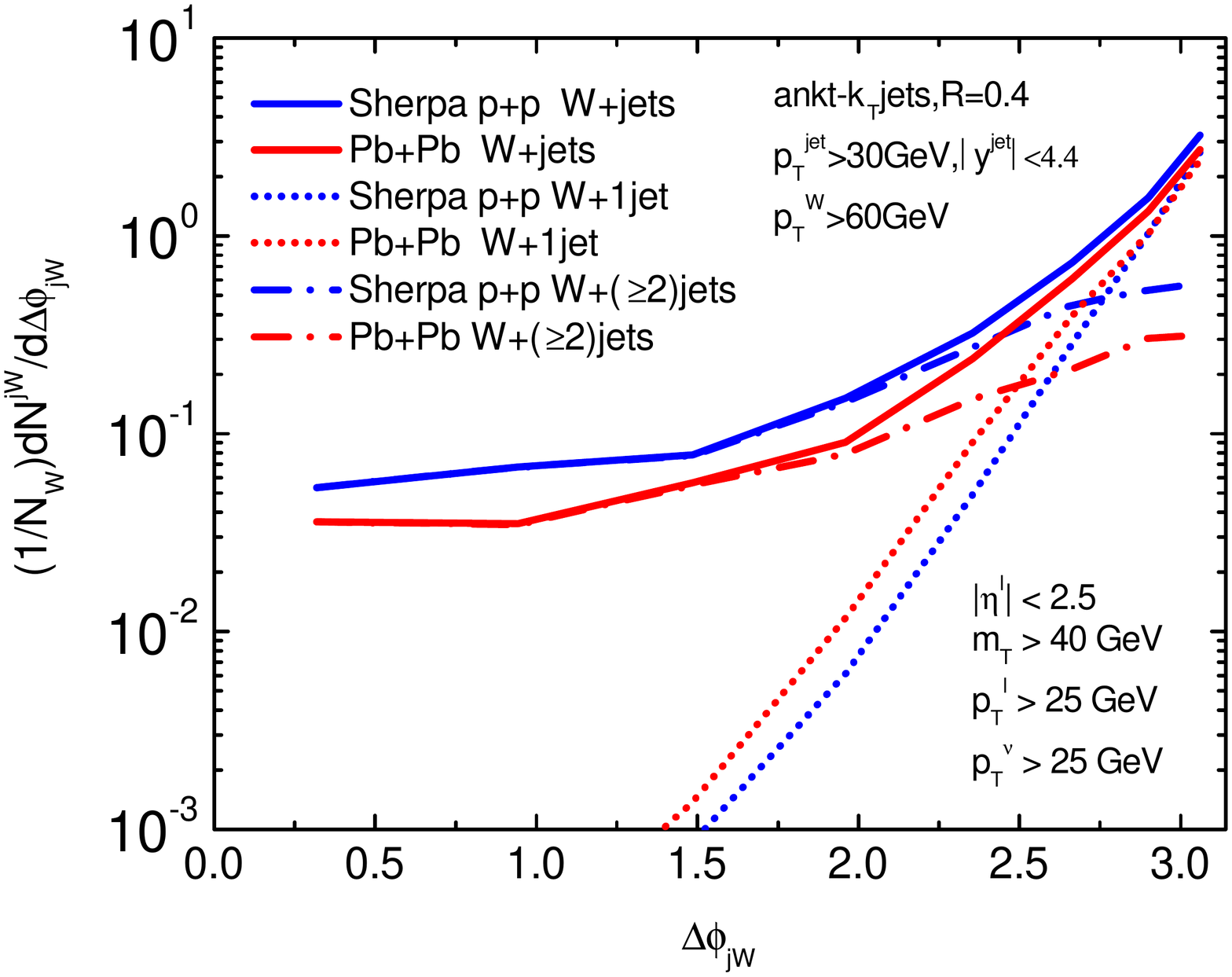}
 \end{center}
   \vspace{-10pt}
  \caption{(Color online) Left: nuclear modification factor $I_{AA}$ in 4 $p_T^W$ intervals. Right: azimuthal correlations $\Delta\phi_{jW}$  of W+jets in central Pb+Pb collisions and p+p collisions at $\sqrt{s_{NN}}=5.02$ TeV.}\label{Iaa_PHI}
\end{figure}

  Fig.~\ref{Iaa_PHI} (left) plots the nuclear modification factor $I_{AA}=(dN^{Pb+Pb}/dp_T^{jet})/(dN^{p+p}/dp_T^{jet})$ of jets tagged by a W within four $p_T^W$ intervals.  An enhancement in $p_T^{jet} < p^{W,cut}_T$ region,  and  a suppression in  $p_T^{jet} > p^{W,cut}_T$ region are observed. We find that $I_{AA}$ is quite sensitive to the kinematic cut due to the steeply falling cross-section in $p_T^W$  region. W+jet azimuthal correlations $\Delta\phi_{jW}=|\phi_{jet}-\phi_W|$  in p+p and Pb+Pb are shown in Fig.~\ref{Iaa_PHI} (right). It is moderately suppressed in Pb+Pb collisions relative to pp collisions. To have a detailed understanding of the suppression, separated contributions from W plus only one jet and W  in association with more than one jet in both p+p and Pb+Pb collisions are also revealed in Fig.~\ref{Iaa_PHI}. We see that W + 1jet dominates in the large angle region and  no significant difference between p+p and Pb+Pb collisions is observed at such high energy scale. However, W+ multi-jet processes dominates in the small angle region and is considerably suppressed in Pb+Pb collisions compared to pp collisions. Because the initial energies of multi-jet are low and  its final energy can easily fall below the thresholds. It is the modification of  W+multi-jets azimuthal angle difference that leads to the suppression of W+jets azimuthal correlations.

\begin{figure}
  \centering
  \includegraphics[scale=0.25]{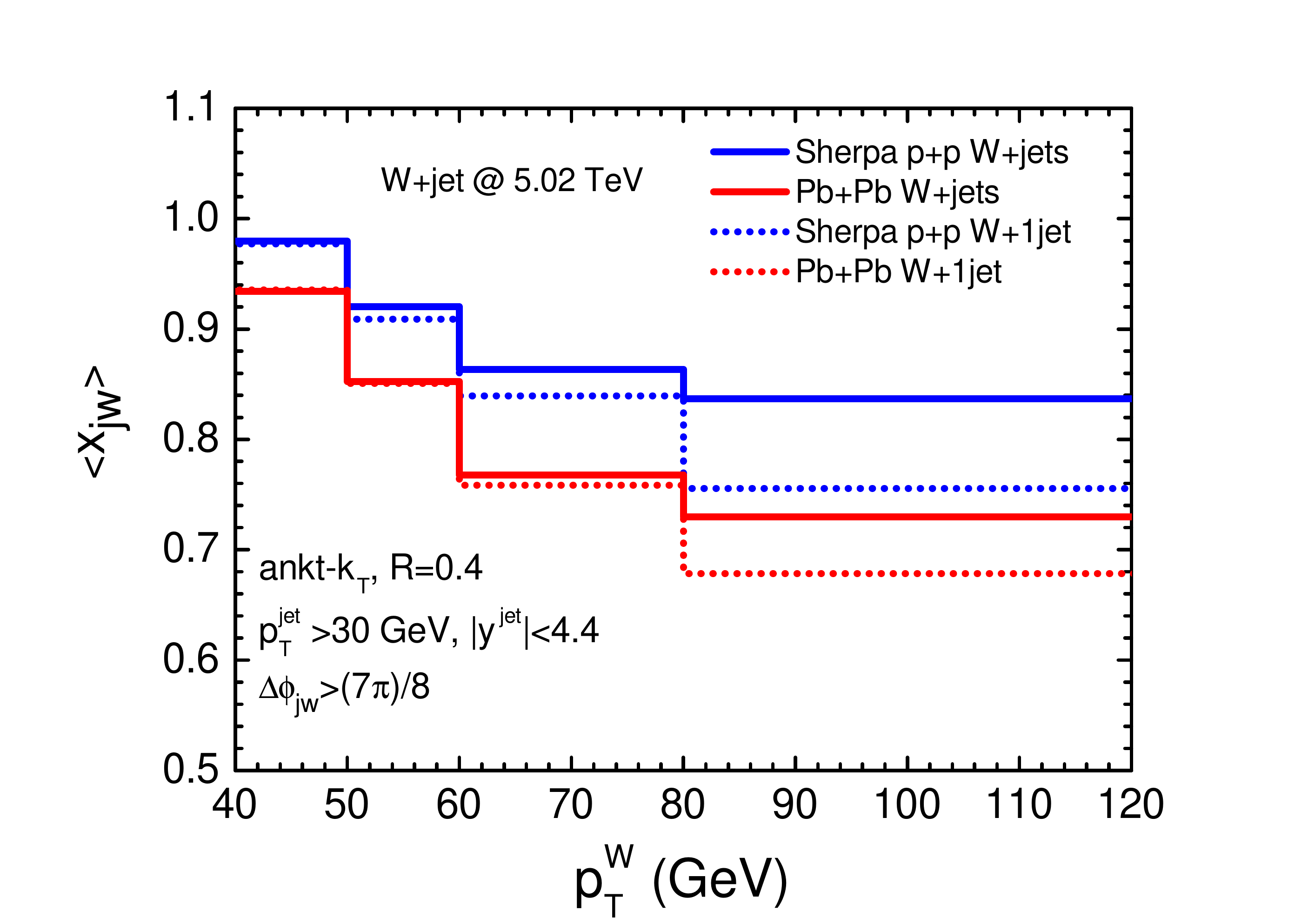}
  \includegraphics[scale=0.25]{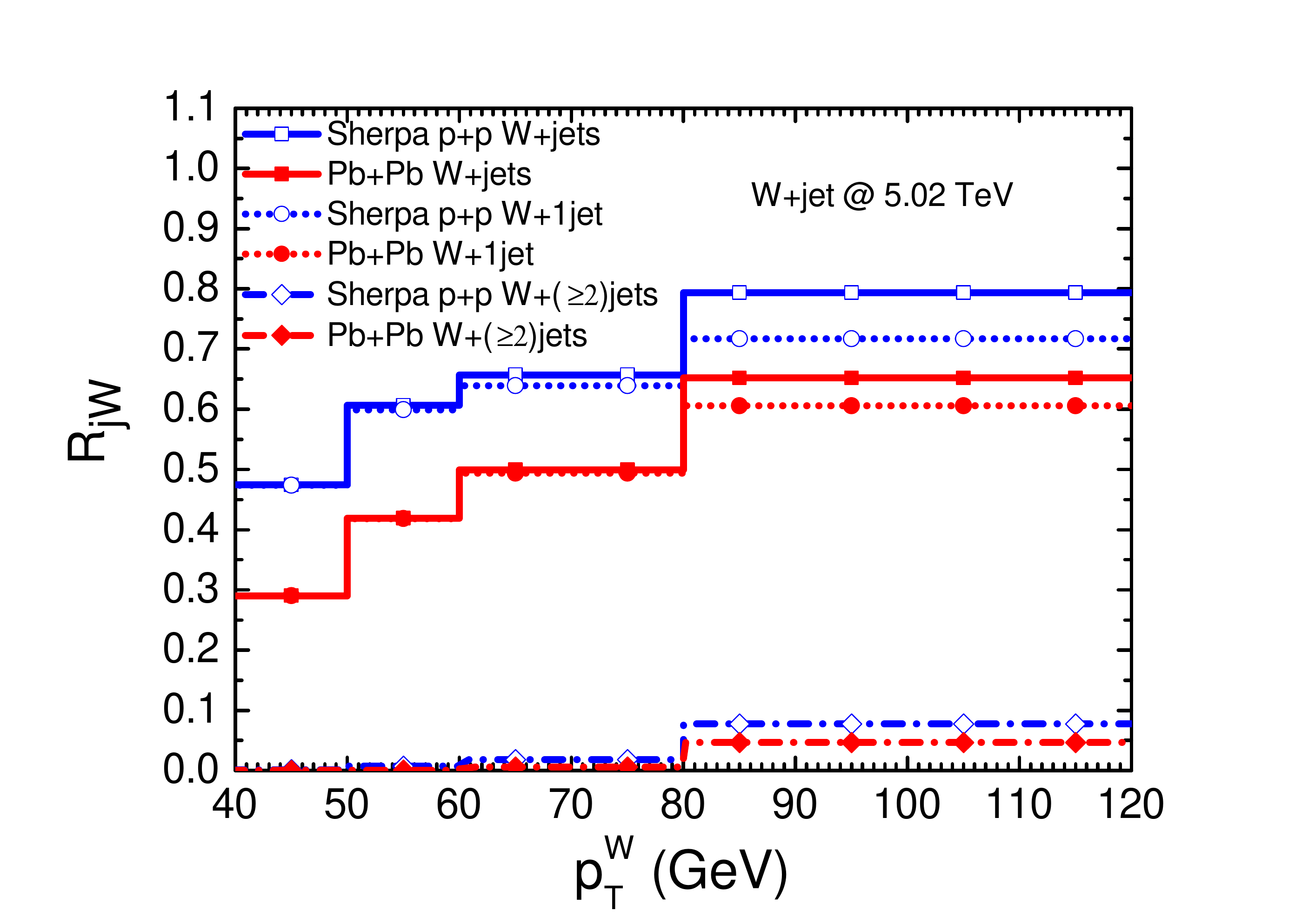}
  \caption{ (Color online) (left) Mean value of momentum imbalance, and (right) average number jet partners per W boson in central Pb+Pb collisions and p+p collisions at $\sqrt{s_{NN}}=5.02$ TeV.}\label{xjz}
\end{figure}

And then, the mean value of  transverse momentum imbalance between the associated jet and the recoiling  W boson  $x_{jW}=p_T^{jet}/p_T^W$  are  presented in Fig.~\ref{xjz} (left). $ \langle x_{jW}\rangle$ in Pb+Pb collisions is much smaller than that in p+p collisions due to jet energy loss in the medium while the transverse momentum of the W boson is unattenuated. The difference of the value between pp and PbPb collisions increases smoothly as a function of the transverse momentum of W boson. It indicates that jets tagged by higher energy W boson lose larger fraction of its energy. Besides, contributions from Multi-jet processes  are essential for high energy W bosons. Average number of jet partners per W boson $R_{jW}$ are shown in Fig.~\ref{xjz} (right). As can be seen,  average number of jet partners per W boson is overall suppressed in Pb+Pb collisions compared to that in p+p collisions, which is a result of the reduction of jet yields that pass the selection cut after jet quenching in Pb+Pb collisions. Besides, higher energy W boson loss smaller fraction of its jet partners.

This research is supported by NSFC of China with Project Nos. 11435004, by NSFC under Grants Nos. 11935007, 11221504 and 11890714, by DOE under Contract No. DE-AC02-05CH11231, and by NSF under Grant No. ACI-1550228.

\end{document}